\documentclass[PAPER, atlasdraft=false,coverpage=false,texlive=2023, UKenglish, cernpreprint, orcidlogo, texmf, orcidlogo]{atlasdoc}
 
\makeatletter
\DeclareOldFontCommand{\rm}{\normalfont\rmfamily}{\mathrm}
\DeclareOldFontCommand{\sf}{\normalfont\sffamily}{\mathsf}
\DeclareOldFontCommand{\tt}{\normalfont\ttfamily}{\mathtt}
\DeclareOldFontCommand{\bf}{\normalfont\bfseries}{\mathbf}
\DeclareOldFontCommand{\it}{\normalfont\itshape}{\mathit}
\DeclareOldFontCommand{\sl}{\normalfont\slshape}{\@nomath\sl}
\DeclareOldFontCommand{\sc}{\normalfont\scshape}{\@nomath\sc}
\makeatother

\usepackage{atlaspackage}
\usepackage[export]{adjustbox}
\usepackage{multirow}
\usepackage{floatrow}
\usepackage{subfig}
\usepackage{booktabs}
 
\usepackage{atlasbiblatex}
 
\usepackage{atlasphysics}
 
\graphicspath{{logos/}{figures/}}
 
\usepackage{ANA-HDBS-2022-03-PAPER-defs}
\usepackage{siunitx}
\usepackage{float}
\floatstyle{plaintop}
\restylefloat{table}

\AtlasTitle{Constraints on the Higgs boson self-coupling  from
single- and double-Higgs production with the ATLAS detector using $pp$ collisions at $\sqrt{s}=13$~TeV}
 
\AtlasVersion{2.1}
 
\AtlasAbstract{
Constraints  on the Higgs boson self-coupling are set by combining
double-Higgs boson analyses in the $b\bar{b}b\bar{b}$, $b\bar{b}\tau^+\tau^-$ and $b\bar{b} \gamma \gamma$
decay channels with single-Higgs boson analyses targeting the $\gamma \gamma$, $ZZ^*$, $WW^*$, $\tau^+ \tau^-$
and $b\bar{b}$ decay channels.
The data used in these analyses were recorded by the ATLAS detector at the LHC in proton--proton collisions
at $\sqrt{s}=13$~TeV and correspond to an integrated luminosity of 126--139~fb$^{-1}$.
The combination of the double-Higgs analyses sets an upper limit of $\mu_{HH} < 2.4$ at 95\% confidence level on the
double-Higgs production cross-section normalised to its Standard Model prediction.
Combining the single-Higgs and double-Higgs analyses, with the assumption that new physics
affects only the Higgs boson self-coupling ($\lambda_{HHH}$), values outside the interval
$-0.4< \kappa_{\lambda}=(\lambda_{HHH}/\lambda_{HHH}^{\textrm{SM}})< 6.3$ are excluded at 95\%
confidence level. The combined single-Higgs and double-Higgs analyses provide results
with fewer assumptions, by adding in the fit more coupling modifiers introduced to
account for the Higgs boson interactions with the other Standard Model particles.
In this relaxed scenario, the constraint becomes $-1.4 <  \kappa_{\lambda} < 6.1$ at 95\% CL.}

\AtlasRefCode{HDBS-2022-03}
 
\PreprintIdNumber{CERN-EP-2022-149}

\AtlasJournal{Phys.\ Lett.\ B.}
\AtlasJournalRef{Phys. Lett. B 843 (2023) 137745}
\AtlasDOI{DOI:10.1016/j.physletb.2023.137745}

\hypersetup{pdftitle={ATLAS document},pdfauthor={The ATLAS Collaboration}}
 
\begin{document}
 
\maketitle

\section{Introduction}
\label{sec:intro}

Since the discovery of the Higgs boson by the ATLAS and CMS collaborations~\cite{HIGG-2012-27, CMS-HIG-12-028} at the Large Hadron Collider
(LHC)~\cite{Evans:2008}, a major goal  of the physics programme of the LHC experiments  has been to measure its properties and determine whether they
correspond to those predicted by the Standard Model (SM) of particle physics~\cite{Glashow:1961tr,Salam:1968rm,Weinberg:1967tq,tHooft:1972tcz}
or involve new phenomena beyond those described by this theory.
One of the most intriguing and interesting characteristics of the SM is that the gauge electroweak (EW) symmetry  is broken spontaneously
by the non-trivial structure of the Higgs boson~\cite{Englert:1964et,Higgs:1964ia,Higgs:1964pj,Higgs:1966ev,Guralnik:1964eu,Kibble:1967sv}
potential, related to its self-interaction.
In the SM, this mechanism allows elementary particles to acquire their mass, while preserving perturbative unitarity up to very high energies.
The Higgs boson potential also plays a fundamental role in understanding the stability of our universe~\cite{Degrassi_2012}.
 
The Higgs boson self-interactions are characterised by the trilinear self-coupling $\lambda_{HHH}$.
In the SM, the Higgs boson self-coupling can be predicted at lowest order from the values of the Higgs boson mass $m_H$~\cite{HIGG-2014-14}
and the Fermi constant $G_\text{F}$~\cite{PDG}: $\lambda_{HHH} = (m_{H}^{2} G_\text{F}) / \sqrt{2}$.
 
At the LHC the Higgs boson self-interaction is directly accessible via the production of Higgs boson
pairs (here referred to as double-Higgs production).
In this Letter the three most sensitive double-Higgs decay channels, $b\bar{b} \gamma \gamma$, $b\bar{b}\tau^+\tau^-$, and $b\bar{b}b\bar{b}$~\cite{HDBS-2018-34,HDBS-2018-40,HDBS-2019-29_temp}, are combined using  the complete dataset collected by ATLAS at $\sqrt{s}=13$~\TeV\  in the data-taking period 2015--2018,
corresponding to an integrated luminosity of 126--139~\ifb.
This combination is used to place constraints on the double-Higgs production cross-section and on the Higgs boson self-coupling.
Results are reported in terms of the coupling modifier \kl\ defined as the ratio of the Higgs boson self-coupling to its
SM value, $\kl = \lambda_{HHH} / \lambda^{\mathrm{SM}}_{HHH}$.
 
The Higgs boson self-interaction also contributes to other processes via sizeable next-to-leading-order (NLO) EW
corrections. In particular, it has been shown~\cite{Degrassi:2016wml,Maltoni:2017ims,DiVita:2017eyz,Gorbahn:2016uoy,Bizon:2016wgr,McCullough:2013rea}  that the single Higgs boson (here referred to as single-Higgs) production cross-sections and branching ratios are also modified if the Higgs boson self-coupling deviates from the SM prediction.
 
More stringent constraints on  $\kl$ are also reported in this Letter from combinations of the recent ATLAS
single-Higgs results~\cite{NaturePaper} based on the full Run 2 data set from the $\gamma \gamma$, $ZZ^*$, $WW^*$, $\tau^+ \tau^-$
and $b\bar{b}$ decay channels with the above mentioned double-Higgs results.
The single-Higgs measurements of the simplified template cross-sections (STXS) and
the double-Higgs results have been parameterised to take into account the impact of \kl\
and the other coupling modifiers. This more comprehensive combination makes it possible to perform tests of \kl\ relaxing
the assumptions about Higgs boson interactions with the other SM particles.
 
A previous ATLAS combination of searches for non-resonant and resonant \HH pair production was performed on a partial Run~2 dataset,
using up to 36.1\,fb$^{-1}$ of data~\cite{HDBS-2018-58}.   The combined observed (expected) upper limit on non-resonant \HH production at 95\% confidence level (CL) was 6.9 (10) times the predicted SM cross-section. When varying the Higgs boson trilinear self-coupling from its SM value, the allowed range of the self-coupling modifier \kl\ was observed (expected) to be $-5.0 \le \kl \le 12.0$ ($-5.8 \le \kl \le 12.0$).
The CMS Collaboration also published a combination of \HH searches using its full Run~2 dataset,
up to 138\,fb$^{-1}$ of data~\cite{CMSNatuerPaper}.   The CMS combined observed (expected) upper limit on non-resonant \HH production at 95\% CL is 3.4 (2.5) times the predicted Standard Model cross-section, and the observed  allowed range of the self-coupling modifier \kl is $-1.24 \le \kl \le 6.49$.


\section{Theoretical framework}
\label{sec:theory}

A simplified way to test the validity of the SM in the Higgs sector is provided by the so called `kappa framework'~\cite{Heinemeyer:2013tqa,deFlorian:2016spz}.
In this framework, the couplings of the Higgs boson to the other SM particles involved at leading order (LO) in perturbation theory for the process
under study are dressed with scaling factors $\kappa_{m}$.
In this simplified approach, based on several assumptions described in Section~10.2 of Ref.~\cite{Heinemeyer:2013tqa}, production and decay yields
are scaled by powers of the corresponding coupling modifier $\kappa_{m}$ defined as the ratio of the coupling between the particle $m$
and the Higgs boson to its SM value.
Any significant deviation of a measured $\kappa_{m}$ from unity would indicate the presence of physics beyond the SM in the tested interaction.
In this work, only the coupling modifiers $\kappa_t$, $\kappa_b$, $\kappa_{\tau}$, and $\kappa_V$ are considered for single-Higgs interactions
(in addition to the \kl\ modifier that impacts the NLO EW corrections as described in the following).
They describe the modifications of the SM Higgs boson coupling to up-type quarks, to down-type quarks, to
leptons and to vector bosons $V$ ($V=W$ or $Z$) respectively.
In this parameterisation the interactions between the Higgs boson and the gluons and photons are
resolved in terms of the coupling modifiers of the SM particles that enter the loop-level diagrams.
New particles contributing to these diagrams are not considered.
The total width of the Higgs boson is also parameterised in terms of the coupling modifiers of the individual SM particles,
assuming no beyond-the-SM contributions.
For double-Higgs production the coupling modifiers \kl, \ktop, \kV and \ktV\ are considered.
The last of these is related to the $VVHH$ interaction vertex, which can be tested in double-Higgs vector-boson fusion (VBF) production (\VBFHH) as described in the following.
 
Double-Higgs production is directly sensitive to the Higgs boson self-coupling, starting at the lowest order in perturbation theory.
In the SM, the gluon--gluon fusion process (\ggFHH) accounts for more than 90\% of the Higgs boson pair-production
$pp \rightarrow HH$ cross-section. The next most abundant process is \VBFHH\ production,
while very small contributions are expected from double-Higgs production in association with a vector boson
($VHH$) and in association with top-quarks ($t\bar{t}\ HH$).
An overview of double-Higgs production at the LHC can be found in Ref.~\cite{hh-whitepaper}.
 
At lowest order in perturbation theory, the  \ggFHH\ process proceeds via two amplitudes: the first ($\mathcal{A}_1$) represented by diagram
(a) in Figure~\ref{fig:hh-diagrams}, and the second ($\mathcal{A}_2$) represented by diagram (b).
The $\mathcal{A}_1$ amplitude is proportional to the square of the Higgs boson coupling to the top-quark, which scales as $\ktop^2$, and the $\mathcal{A}_2$
amplitude is proportional to the product of \ktop\  and the Higgs boson self-coupling modifier \kl.
\begin{figure}[htbp]
\centering
\subfloat[]{\includegraphics[width=0.35\textwidth,valign=c]{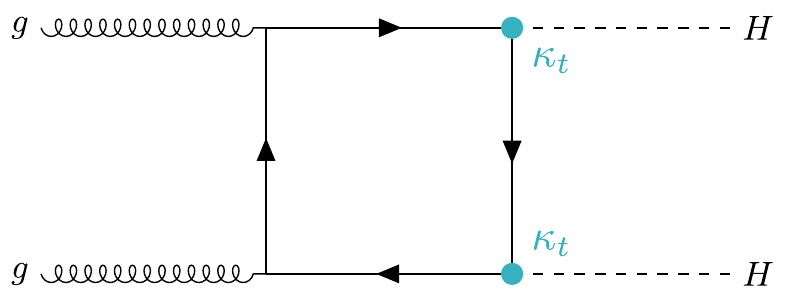}\label{fig:hh-diagrams_a}}
\subfloat[]{\includegraphics[width=0.35\textwidth,valign=c]{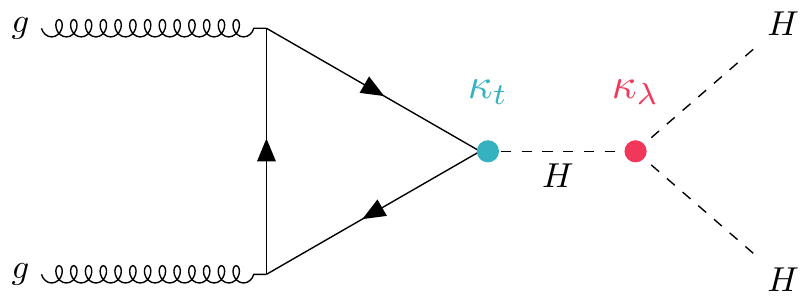}\label{fig:hh-diagrams_b}} \\
\quad \quad
\subfloat[]{\includegraphics[width=0.25\textwidth,valign=c]{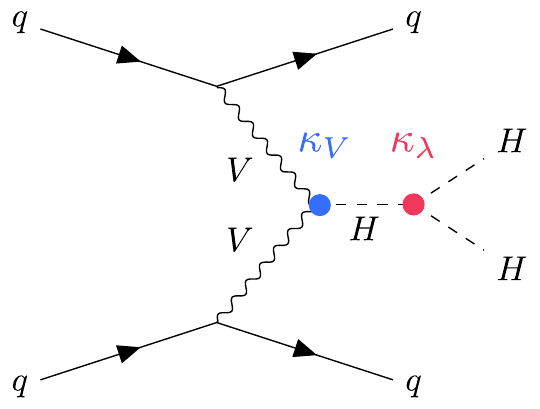}\label{fig:hh-diagrams_c}}
\subfloat[]{\includegraphics[width=0.25\textwidth,valign=c]{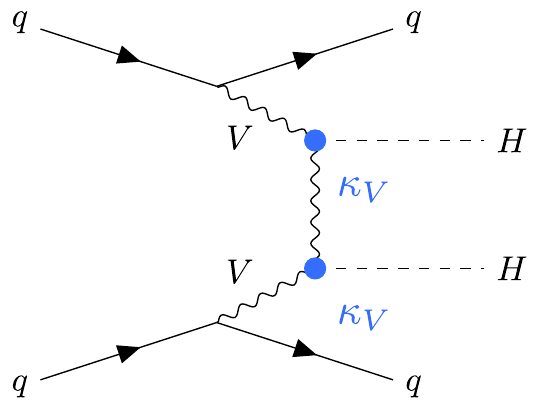}\label{fig:hh-diagrams_d}}
\subfloat[]{\includegraphics[width=0.25\textwidth,valign=c]{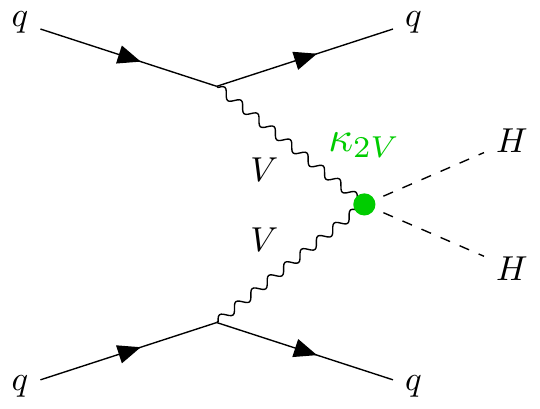}\label{fig:hh-diagrams_e}}
 
\quad \quad
\caption{Examples of leading-order Feynman
diagrams for Higgs boson pair production: for ggF production, diagram
(a) is proportional to the square of the top-quark
Yukawa coupling, while diagram (b) is proportional to the product
of the top-quark Yukawa coupling and the Higgs boson self-coupling.
For VBF production, diagram (c) is proportional to the product of the coupling of the Higgs boson to the vector bosons
and the self-coupling,
diagram (d) to the square of the coupling to the vector bosons, and
diagram (e) to the interaction between two vectors bosons and two Higgs bosons.}
\label{fig:hh-diagrams}
\end{figure}
 
In the SM, the interference between these two amplitudes is destructive and
yields an overall cross-section of $\sigma^{\mathrm{SM}}_{\mathrm{ggF}} (pp \to HH) = 31.0^{+2.1}_{-7.2}$~fb at
$\sqrt{s} = 13$~\TeV, calculated at NLO in QCD with the measured value of the top-quark mass and corrected to next-to-next-to-leading order (NNLO)
including finite top-quark mass effects~\cite{Grazzini:2018bsd, Heinrich:2019bkc, deFlorian:2016spz,Dawson:1998py,Borowka:2016ehy,Baglio:2018lrj, Bonciani:2018omm, deFlorian:2013jea,Shao:2013bz,deFlorian:2015moa,Baglio:2021}. The large negative uncertainty originates from the scheme and scale choice of the virtual top-quark mass~\cite{Baglio:2021}.
Deviations of the \ggFHH\ cross-section from the SM prediction can therefore be parameterised in terms of the two coupling
modifiers \kl\ and \ktop\ following the prescription described in
Refs.~\cite{deFlorian:2016spz,Dawson:1998py,Borowka:2016ehy, Baglio:2018lrj, Bonciani:2018omm, deFlorian:2013jea,Shao:2013bz,deFlorian:2015moa}.
Higher-order QCD corrections do not add further \ttH\  or $HHH$ vertices to the diagrams shown in
Figure \ref{fig:hh-diagrams}, implying that this parameterisation is applicable to any order
in QCD (i.e.\ also when the amplitudes $\mathcal{A}_1$ and $\mathcal{A}_2$ are modified to include their
higher-order QCD corrections).
Signal samples for \ggF\ double-Higgs production can be obtained from simulated samples that are generated at different values of these couplings and
then combined using morphing techniques, as described in Ref.~\cite{HDBS-2018-58}. Detailed validation studies of this procedure can be
found in Ref.~\cite{ATL-PHYS-PUB-2019-007}.
In the SM, the $b$-quark loop contribution to the \ggFHH\ cross-section is negligible~\cite{Baglio:2020,deFlorian:2016spz,Dawson:2012mk,Plehn:1996wb},
so its contribution is not included in this analysis.
 
The second most abundant SM double-Higgs process is \VBFHH\ production, with a predicted SM cross-section
of $1.72 \pm 0.04$~fb at 13 TeV~\cite{Dreyer:2018,Baglio:2013,Ling:2014}.
At LO in perturbation theory, this process depends on several diagrams
that involve the interaction of the Higgs boson with the $W$ or $Z$ vector bosons as shown in
Figure~\ref{fig:hh-diagrams}.
The three representative diagrams that enter the total amplitude of the \VBFHH\ process can be parameterised with different combinations
of the \kl, $\kappa_{V}$ and \ktV\ coupling modifiers~\cite{Bishara:2016kjn}. The first diagram, shown in Figure~\ref{fig:hh-diagrams_c}, is proportional to \kV\ and \kl,
the second, shown in Figure~\ref{fig:hh-diagrams_d}, to $\kV^{2}$ and the last one, shown in Figure~\ref{fig:hh-diagrams_e} and
related to the quartic interaction vertex $VVHH$, to \ktV.
The \VBFHH\ production process can therefore be parameterised using six terms derived from the square of
the amplitude described above, which scales as a polynomial of \kl, $\kappa_{V}$ and \ktV.
The parameterisation of the signal samples, in terms of yields and kinematic properties, for the double-Higgs VBF process as a function of these coupling modifiers is performed using
a set of six independent samples generated  for different values of \kl, \kV\ and \ktV.
The values of \kl, \kV, and \ktV\ for these six samples were chosen to obtain good statistical precision in the region of parameter space where this analysis is sensitive.
The validity of this parameterisation was checked with additional VBF signal samples generated with different values of these coupling modifiers.
 
The \ggFHH\ process is sensitive to the sign of \kl\ relative to the top-quark couplings because of
interference between different amplitudes whose leading-order Feynman diagrams are depicted in Figure~\ref{fig:hh-diagrams}.
Similarly, the \VBFHH\ process provides sensitivity to the relative sign between \ktV\ and \kV.

A complementary approach to study the Higgs boson self-coupling is to use single-Higgs processes, as proposed in
Refs.~\cite{Degrassi:2016wml,Maltoni:2017ims,DiVita:2017eyz,Gorbahn:2016uoy,Bizon:2016wgr,McCullough:2013rea}.
These processes do not depend on $\lambda_{HHH}$ at LO, but the Higgs boson self-coupling
contributes to the calculation of the complete NLO EW corrections.
In particular, $\lambda_{HHH}$ contributes to NLO EW corrections via Higgs boson self-energy loop corrections and via additional diagrams, examples of which are
shown in Figure~\ref{fig:self_coupling_diagram}.
\begin{figure}[tbp]
\centering
\subfloat[]{\includegraphics[width=0.32\textwidth]{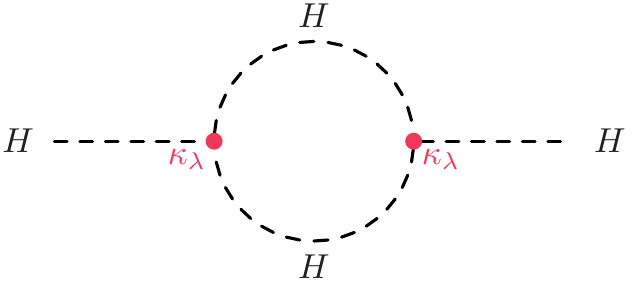}\label{fig:diag_self}} \\
\subfloat[]{\includegraphics[width=0.32\textwidth,valign=c]{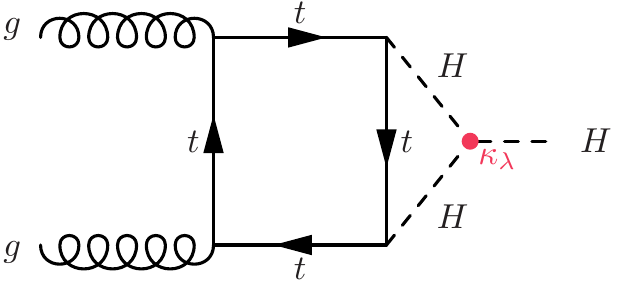}}
\subfloat[]{\includegraphics[width=0.32\textwidth,valign=c]{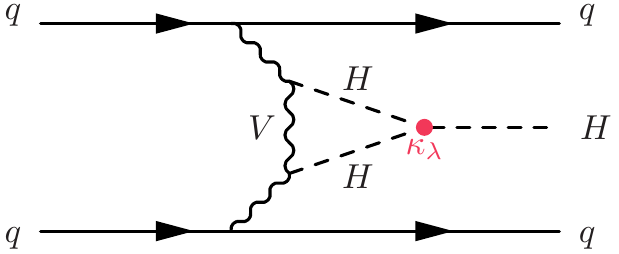}}
\hfill
\subfloat[]{\includegraphics[width=0.32\textwidth,valign=c]{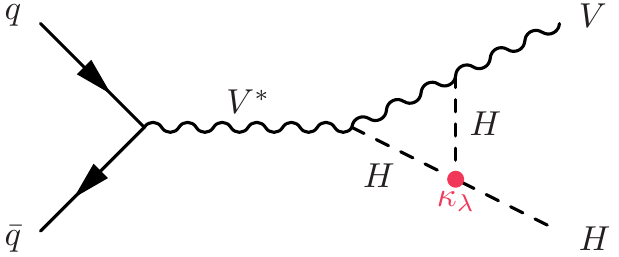}}
\hfill
\subfloat[]{\includegraphics[width=0.32\textwidth]{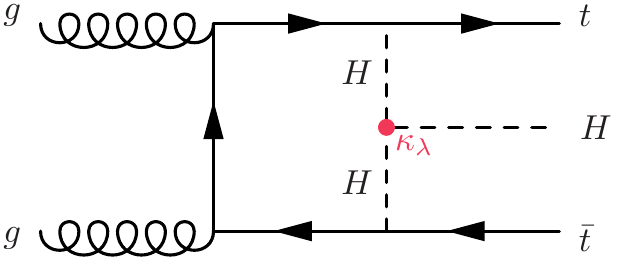}}
\caption{Examples of one-loop $\lambda_{HHH}$-dependent diagrams for (a) the Higgs boson self-energy, and for
single-Higgs production in the (b) \ggF, (c) \VBF,  (d) \VH,  and (e) \ttH\ modes. The self-coupling vertex is indicated by the filled circle.  }
\label{fig:self_coupling_diagram}
\end{figure}
Therefore, an indirect constraint on \kl\ can be extracted by comparing precise measurements
of single-Higgs production and decay yields with the SM predictions corrected for the $\lambda_{HHH}$-dependent NLO EW effects.
A framework for a global fit to constrain the Higgs boson self-coupling and the other coupling modifiers $\kappa_m$
was proposed in Refs.~\cite{Degrassi:2016wml,Maltoni:2017ims}; the model-dependent assumptions of this parameterisation
are described in the same references.
In the current work, inclusive production cross-sections, decay branching ratios and differential cross-sections are exploited to increase the
sensitivity of the single-Higgs analyses to $\kl$ and $\kappa_m$.
The differential information is encoded through the simplified template cross-section (STXS) framework described in Section~III.3 of
Ref.~\cite{Badger:2016bpw}. 
The signal yield in a specific decay channel and STXS bin is then proportional to:
\begin{equation*}
n_{i,f}^{\text{signal}}(\kl, \kappa_m) \propto  \mu_{i}(\kappa_{\lambda}, \kappa_m) \times \mu_{f}(\kappa_{\lambda}, \kappa_m) \times \sigma_{\textrm{SM},i} \times \mathcal{B}_{\textrm{SM},f} \times (\epsilon \times A)_{if}\,,
\end{equation*}
 
where $\mu_i$  and $\mu_f$ describe respectively the multiplicative corrections to the expected SM Higgs boson production cross-sections in an STXS bin
($\sigma_{\text{SM},i}$) and each decay-channel branching ratio ($\mathcal{B}_{\text{SM},f}$) as a function of the values of the Higgs boson self-coupling
modifier $\kl$ and the LO-inspired modifiers $\kappa_m$.
The $(\epsilon \times A)_{if}$ coefficients take into account the analysis efficiency times acceptance in each production and decay mode.
 
The functional dependence of $\mu_i(\kl,\kappa_m)$ and $\mu_f(\kl,\kappa_m)$ on $\kl$ and $\kappa_m$ varies according to the production mode,
the decay channel and, more strongly  for the  \VH\  and \ttH\ production modes, on the STXS bin.
A detailed description of the cross-section and decay-rate dependence on \kl\ is given in Refs.~\cite{ATL-PHYS-PUB-2019-009,LHCHXSWG-Pubnote}.
The STXS information from the \VBF , \WH , \ZH\ and \ttH\ production modes is exploited here to constrain $\kl$ and $\kappa_m$.
For the ggF production mode, only the inclusive cross-section dependence on $\kl$ is currently available and it was
used in this study, while the STXS bin dependence was not considered.
 
Conversely, the $\kl$-modifier can affect the Higgs boson production kinematics and thus modify the analysis efficiency times acceptance in a given STXS bin.
This residual dependence was evaluated and found to be negligible for single-Higgs processes, as described in Ref.~\cite{ATL-PHYS-PUB-2019-009}.
Thus the single-Higgs selection acceptances and efficiencies are assumed to be constant as a function of \kl\ in each STXS bin.
A detailed description of the parameterisation of the single-Higgs processes as a function of the \kl\ coupling modifiers
used in this Letter can be found in Ref.~\cite{LHCHXSWG-Pubnote}.
The model under discussion does not allow for any new physics beyond that encoded in the aforementioned $\kl$ and $\kappa_m$ parameters.
The dependence of the decay branching ratios and the Higgs boson self-energy on \kl\ is also taken into account for the double-Higgs analyses when combining them with the single-Higgs results.
 
A Higgs boson mass value of $m_{H} = 125.09 \pm 0.24$~\GeV~\cite{HIGG-2014-14} is used for all results presented in this Letter.


\section{Data samples and combined analyses}
\label{sec:ana_descr}

The results, presented in Sections~\ref{sec:hhcombo} and~\ref{sec:hhplushcombo}, are obtained using the full Run~2 dataset collected by the ATLAS experiment~\cite{PERF-2007-01,Capeans:1291633, Abbott:2018ikt}
from LHC 13~\TeV\ $pp$ collisions in the 2015--2018 data-taking period.
The integrated luminosity corresponds to 126--\currentluminosity , depending on the trigger selection.
A two-level trigger system~\cite{TRIG-2016-01} is used to select events.
An extensive software suite~\cite{ATL-SOFT-PUB-2021-001} is used in the reconstruction and analysis of collision
and simulated data, in detector operations, and in the trigger and data acquisition systems of the experiment.
 
Each input analysis used in the combination is summarised in Table~\ref{tab:analysis}. Details about the individual
analyses can be found in the references reported in the same table.
Each analysis separates the selected events into different kinematic and topological regions, called categories.
 
\begin{table}[!htbp]
\caption{Integrated luminosity of the dataset used for each input
channel in the combination. The last column provides references to
publications describing each channel in detail.}
\begin{center}
\begin{tabular}{lcc}
\toprule
Analysis channel & Integrated luminosity [fb$^{-1}$] & Ref. \\
\midrule
\HHbbyy & 139 & \cite{HDBS-2018-34} \\
\HHbbtt & 139 & \cite{HDBS-2018-40}\\
\HHbbbb & 126 & \cite{HDBS-2019-29_temp}\\
\midrule
\hyy      & 139 & \cite{ATLAS:2022tnm} \\
\hZZllll  & 139 & \cite{HIGG-2018-28} \\
\htt      & 139 & \cite{HIGG-2019-09}\\
\hww~~(\tggF ,\tVBF)      & 139 & \cite{ATLAS:2022ooq} \\
\hbb~~~~(\tVH)                  & 139 & \cite{HIGG-2018-51} \\
\hbb~~~~(\tVBF)                 & 126 & \cite{HIGG-2019-04} \\
\hbb~~~~(\tttH)                 & 139  & \cite{HIGG-2020-23} \\
\bottomrule
\end{tabular}
\end{center}
\label{tab:analysis}
\end{table}


\section{Statistical model  and systematic uncertainty correlations}
\label{sec:statmodel}

The statistical treatment used in this Letter follows the procedures described in Refs.~\cite{HIGG-2015-07,HIGG-2018-57}.
The results are obtained from a likelihood function $L(\vec{\alpha},\vec{\theta})$, where $\vec{\alpha}$ represents the vector of the parameters of interest (POI) of the model and $\vec{\theta}$ is a set of nuisance parameters, including the systematic uncertainty
contributions and background parameters that are  constrained by sidebands or control regions in data.
The global likelihood function $L(\vec{\alpha},\vec{\theta})$ is obtained as the product of the likelihoods of each input analysis.
These are, in turn, products of likelihoods computed in the single analysis categories.
The results presented in the following sections are based on the profile-likelihood-ratio test statistic $\Lambda(\vec{\alpha},\vec{\theta})$,
and 68\% as well as 95\% CL intervals are derived in the asymptotic approximation~\cite{Cowan:2010js}.
The $\mathrm{CL_{s}}$ approach~\cite{Read:2002hq} is only used to derive the cross-section upper limits shown in Section~\ref{sec:hhcombo}.
 
To derive the expected results, Asimov datasets~\cite{Cowan:2010js} are produced with all the nuisance parameters set to the values derived from
the fit to the data and the parameters of interest fixed to the values corresponding to the hypothesis mentioned in the text.
 
The basic assumption in performing a statistical combination by using the product of the likelihoods is that the analyses being combined are
statistically independent. For this reason the event samples used in the single-Higgs and double-Higgs
analyses were checked for overlaps. The overlap among the single-Higgs analyses was checked previously in the combination published in
Ref.~\cite{NaturePaper} and found to be negligible.
The event overlap among the three double-Higgs analyses combined for the first time for this result was studied and found to be significantly smaller than 0.1\%.
These analyses are therefore treated as statistically independent.
As a last step, the overlap of event samples between the single-Higgs and double-Higgs analyses,
which are combined for the first time in this Letter, was investigated.
For most of the categories, this overlap is significantly below the 1\% level in either the single-Higgs or the double-Higgs channel,
and can therefore be neglected.
The only exception is the overlap between the $H \to \tau^+\tau^-$ and $HH \to b\bar{b}\tau^+\tau^-$ channels, mainly due to
the \ttH\ categories in the $H \to \tau^+\tau^-$ analysis, which is found to be at the 4\% level in the double-Higgs signal regions.
The  \ttH\ categories in the $H \to \tau^+\tau^-$ channel were removed from the combination
used to produce the results presented in the following sections.

A complete discussion of the sources of systematic uncertainty considered in the individual analyses is provided in the
publications referenced in Table~\ref{tab:analysis}.
The correlation model adopted for the systematic uncertainties within the single-Higgs combination is described
in detail in Ref.~\cite{NaturePaper}.
 
For this Letter, additional correlations of systematic uncertainties between the double-Higgs analyses and between the single-Higgs and double-Higgs combinations
were investigated and implemented as needed.
In both cases, systematic uncertainties related to the data-taking conditions, such as those associated with pile-up mis-modelling and the integrated luminosity,
are considered to be fully correlated among the input searches.
Uncertainties related to physics objects used by multiple searches are treated as correlated where appropriate: experimental uncertainties
that are related to the same physics object but determined with different methodologies or implemented with different parameterisations are treated as uncorrelated.
Theoretical uncertainties of simulated signal and background processes, such as the single-Higgs and double-Higgs production cross-sections, QCD scale, and proton parton distribution functions are treated as correlated where relevant. The experimental uncertainty of the Higgs boson mass measurement~\cite{HIGG-2014-14} is treated as correlated where relevant.
Signal theory uncertainties of the single-Higgs and double-Higgs production modes (e.g., missing higher-order QCD corrections, parton shower, parton distribution functions, etc.)  are treated as uncorrelated, while the systematic uncertainties of the decay branching ratios are treated as correlated.
For the systematic uncertainties that are constrained significantly in the fit to data, the impact of treating them as correlated or uncorrelated in the combined fit
was checked. In general, the impact of these different correlation schemes on the exclusion limits is found to be very small, below the 2\% level.
Since choosing to treat them as uncorrelated gives slightly larger uncertainties for the parameter of interest,
this approach was chosen for the results presented in the following sections.
 
For the double-Higgs analyses, the most important uncertainties are related to background estimates from data-driven methodologies (derived from data sidebands
or control regions) and are therefore not correlated with the single-Higgs analyses.
The change of the correlation scheme was found to have a negligible impact on the combined double-Higgs results, except
for the theoretical uncertainties of the \ggFHH\ cross-section,
where assuming a correlation loosens the limits on the signal strength by 7\% and this is therefore adopted.


\section{Double-Higgs combination results}
\label{sec:hhcombo}

The double-Higgs boson analyses in the
$b\bar{b}b\bar{b}$, $b\bar{b}\tau^+\tau^-$ and $b\bar{b} \gamma \gamma$ decay
channels referenced in Table~\ref{tab:analysis} are combined in order to place constraints on the production cross-section and
the Higgs boson's self-coupling.
First, the value of the signal strength $\mu_{HH}$,
defined as the ratio of the double-Higgs production cross-section, including only the \ggFHH\ and \VBFHH\ processes, to its SM prediction
of 32.7~fb~\cite{Grazzini:2018bsd, Heinrich:2019bkc, hh-whitepaper,deFlorian:2016spz,Dawson:1998py,Borowka:2016ehy,Baglio:2018lrj, Bonciani:2018omm, deFlorian:2013jea,Shao:2013bz,deFlorian:2015moa,Dreyer:2018} is determined.
To produce this result the ratio of the \ggFHH\ to \VBFHH\ production cross-sections and the relative kinematic distributions are assumed to be as predicted by the SM, and the other minor production modes are neglected.
 
This combination yields an observed 95\% CL upper limit on $\mu_{HH}$ of 2.4, with an expected upper limit of
2.9 in the absence of $HH$  production and 4.0 expected in the SM case.
The limits on the signal strength obtained from the individual channels and their combination are shown in Figure~\ref{fig:hh-xs-exclusion-channels}. The best-fit value obtained from the fit to the data is $\mu_{HH} = -0.7 \pm 1.3$, which is compatible with the SM prediction of unity, with a $p$-value of 0.2.
\begin{figure}[htbp]
\centering
\includegraphics[width=0.5\textwidth,valign=c]{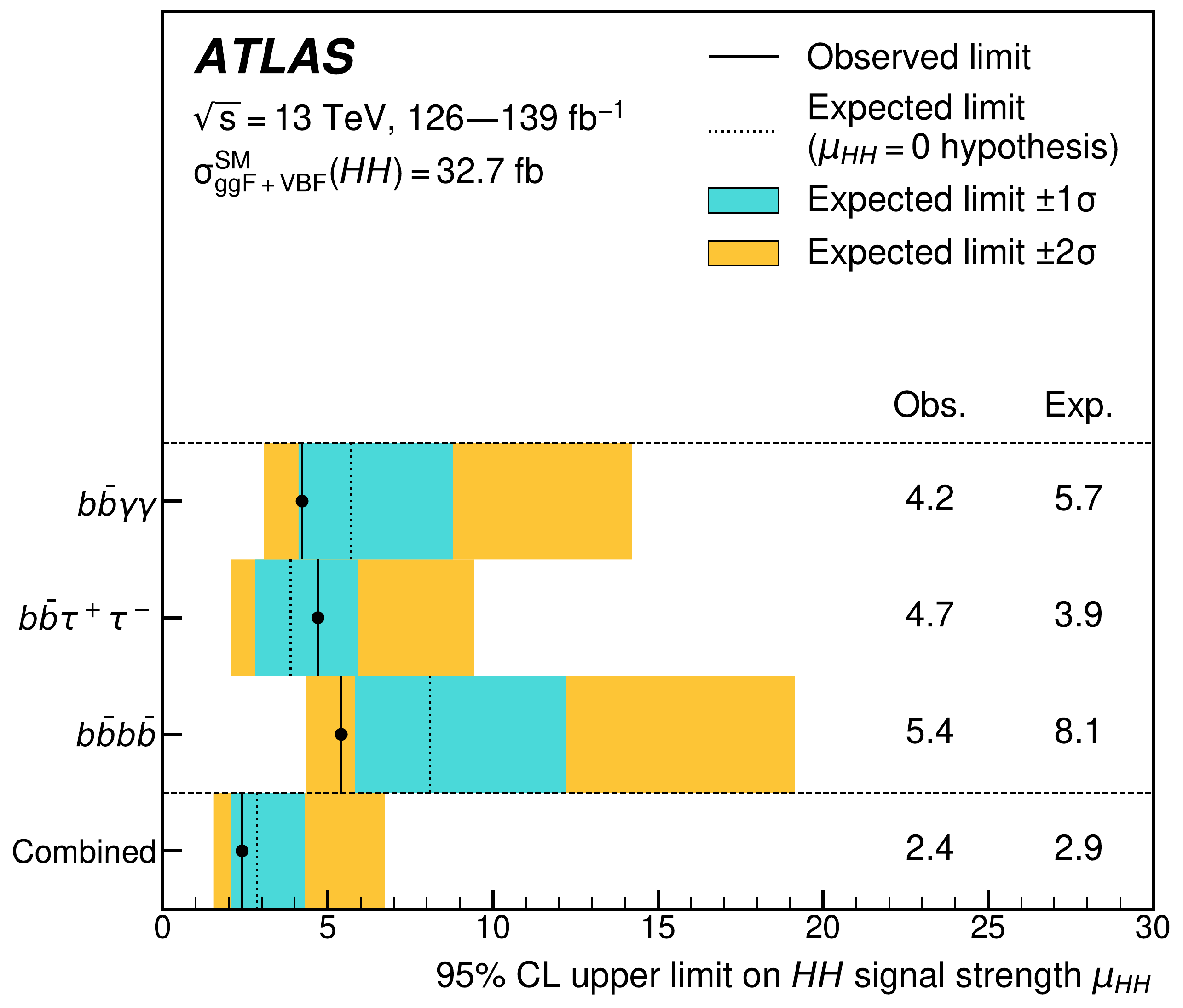}
 
\caption{Observed and expected 95\% CL upper limits on the signal strength
for double-Higgs production from the $b\bar{b}b\bar{b}$, $b\bar{b}\tau^+\tau^-$ and $b\bar{b} \gamma \gamma$ decay
channels, and their statistical combination. The value $m_H = 125.09$ \GeV\ is assumed when deriving the predicted
SM cross-section. The expected limit and the corresponding error bands are derived assuming the absence of the $HH$ process and
with all nuisance parameters profiled to the observed data.}
\label{fig:hh-xs-exclusion-channels}
\end{figure}
From the same combination, a 95\% CL upper limit on $\sigma(pp \to HH)$ of 73~fb is derived
(where only \ggFHH\ and \VBFHH\ processes are considered), compared with an expected limit of 85~fb assuming no $HH$ production.
When deriving the cross-section limits the theoretical uncertainties on the predicted cross-sections
are not included.
The cross-section limit as a function of the coupling modifier is shown in Figure~\ref{fig:hh-xs-exclusion_a}.
The signal acceptance of the double-Higgs analyses has a strong dependence on the value of \kl\ (mainly due to its impact on the $m_{HH}$ distribution), determining the
shapes of the exclusion limit curve shown in  Figure~\ref{fig:hh-xs-exclusion_a}.

\begin{figure}[htbp]
\centering
\subfloat[]{\includegraphics[width=0.45\textwidth,valign=c]{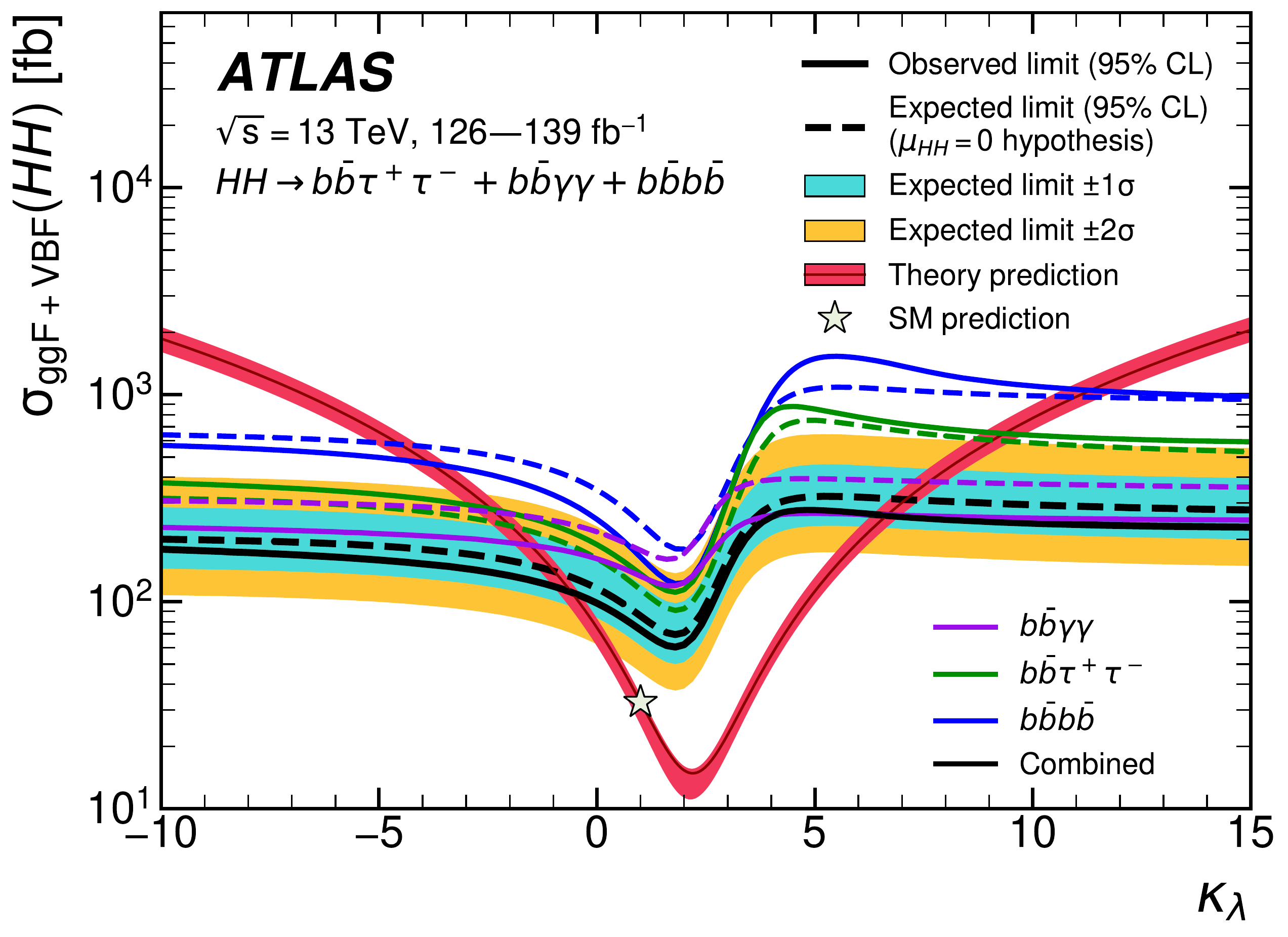}\label{fig:hh-xs-exclusion_a}}
\subfloat[]{\includegraphics[width=0.45\textwidth,valign=c]{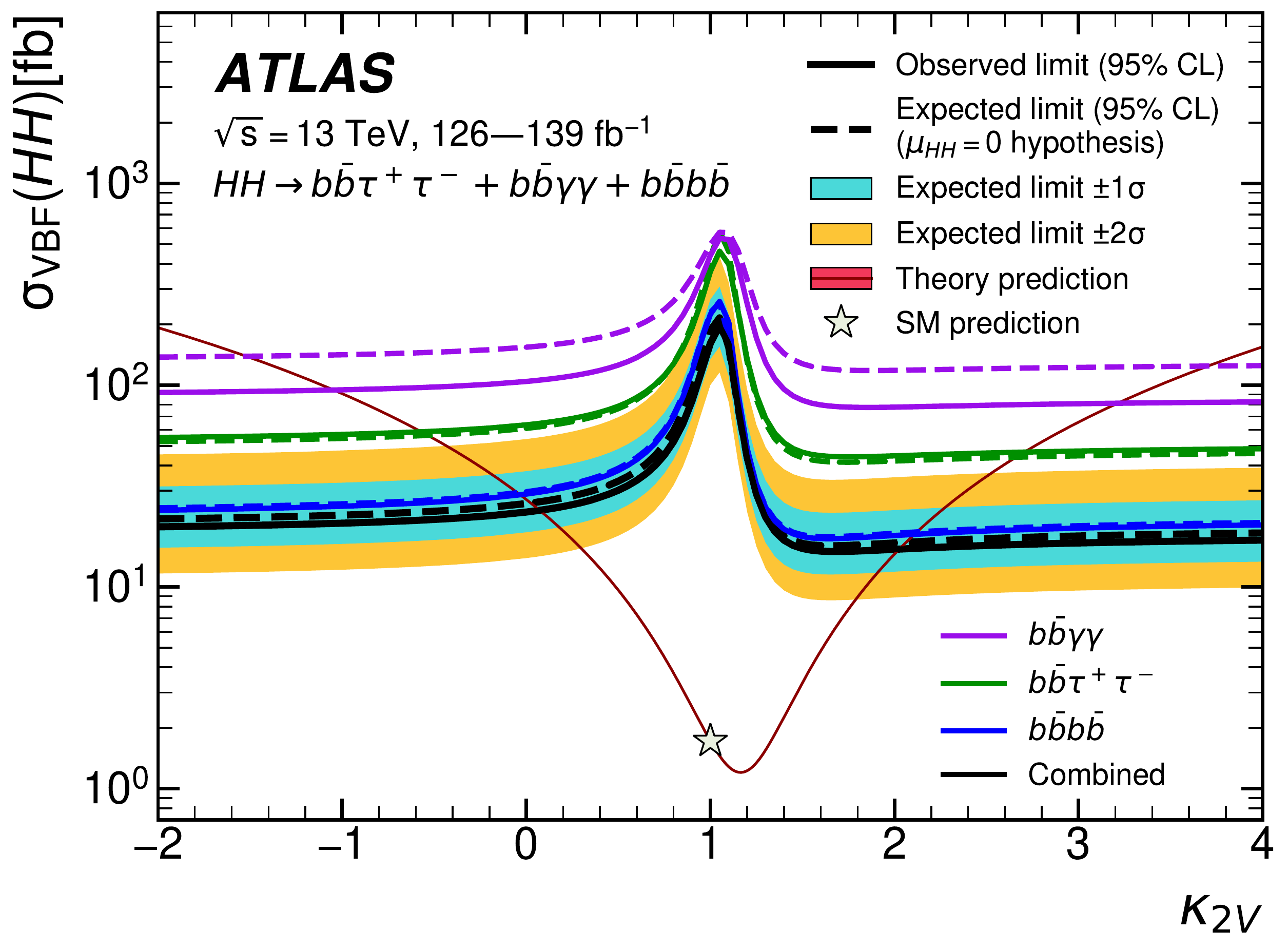}\label{fig:hh-xs-exclusion_b}}
\caption{Observed and expected 95\% CL exclusion limits on the production cross-sections of (a) the combined \ggFHH\ and \VBFHH\ processes as a function of \kl\ and
(b) the \VBFHH\ process as a function of \ktV, for the three double-Higgs search channels and their combination.
The expected limits assume no $HH$ production or no \VBFHH\ production respectively.
The red line shows (a) the theory prediction  for the combined \ggFHH\ and \VBFHH\  cross-section as a function of \kl where all parameters and couplings are set
to their SM values except for \kl ,  and (b) the predicted \VBFHH\ cross-section as a function of \ktV. The  bands surrounding the red cross-section lines indicate
the theoretical uncertainty on the predicted cross-section. The uncertainty band in (b) is smaller than the width of the plotted line.}
\label{fig:hh-xs-exclusion}
\end{figure}

Constraints on the coupling modifiers are obtained by using the values of the test statistic as a function of \kl\ in the
asymptotic approximation and including the theoretical uncertainty of the cross-section predictions.
The \kl\ parameterisation of NLO EW corrections in the Higgs boson decay and self-energy, as well as in single-Higgs backgrounds,
is included when deriving these results, although its impact on the constraints is negligible.
With these assumptions, the observed (expected) constraints at 95\% CL are $-0.6 < \kl < 6.6$ ($-2.1< \kl <7.8$).
The expected constraint is derived using the SM assumption.
More results with different assumptions about the other coupling modifiers are given in Section~\ref{sec:hhplushcombo}.
 
The combined double-Higgs channels are also sensitive to the \VBFHH\ process, and hence to the $HHVV$ quartic interaction.
The 95\% CL observed \VBFHH\ cross-section upper limit as a function of \ktV\ is shown in Figure~\ref{fig:hh-xs-exclusion_b}.
Constraints are derived directly from the test statistic value parameterised as a function of \ktV.
An observed (expected) 95\% CL constraint of $0.1 < \ktV < 2.0$ ($0.0 < \ktV < 2.1$) is obtained,
fixing all other coupling modifiers to unity and with the expected values derived under the SM hypothesis.


\section{Single- and double-Higgs combination results}
\label{sec:hhplushcombo}

Following the prescriptions described in Section~\ref{sec:theory} the double-Higgs and single-Higgs analyses summarised
in Table~\ref{tab:analysis} are combined to derive constraints on \kl.
Several fits to data are performed with different assumptions about the coupling modifiers to other SM particles.
 
At first, only possible deviations of \kl\ from its SM value are considered, assuming that all other Higgs boson interactions
proceed as predicted by the SM. The values of twice the negative-logarithm of the profile likelihood ratio ($-2 \ln{\Lambda}$) as a function of
\kl are shown in Figure~\ref{fig:h-hh-klambda-scan} for the single-Higgs and double-Higgs analyses, and their combination.
\begin{figure}[htbp]
\centering
\subfloat[]{\includegraphics[width=0.5\textwidth,valign=c]{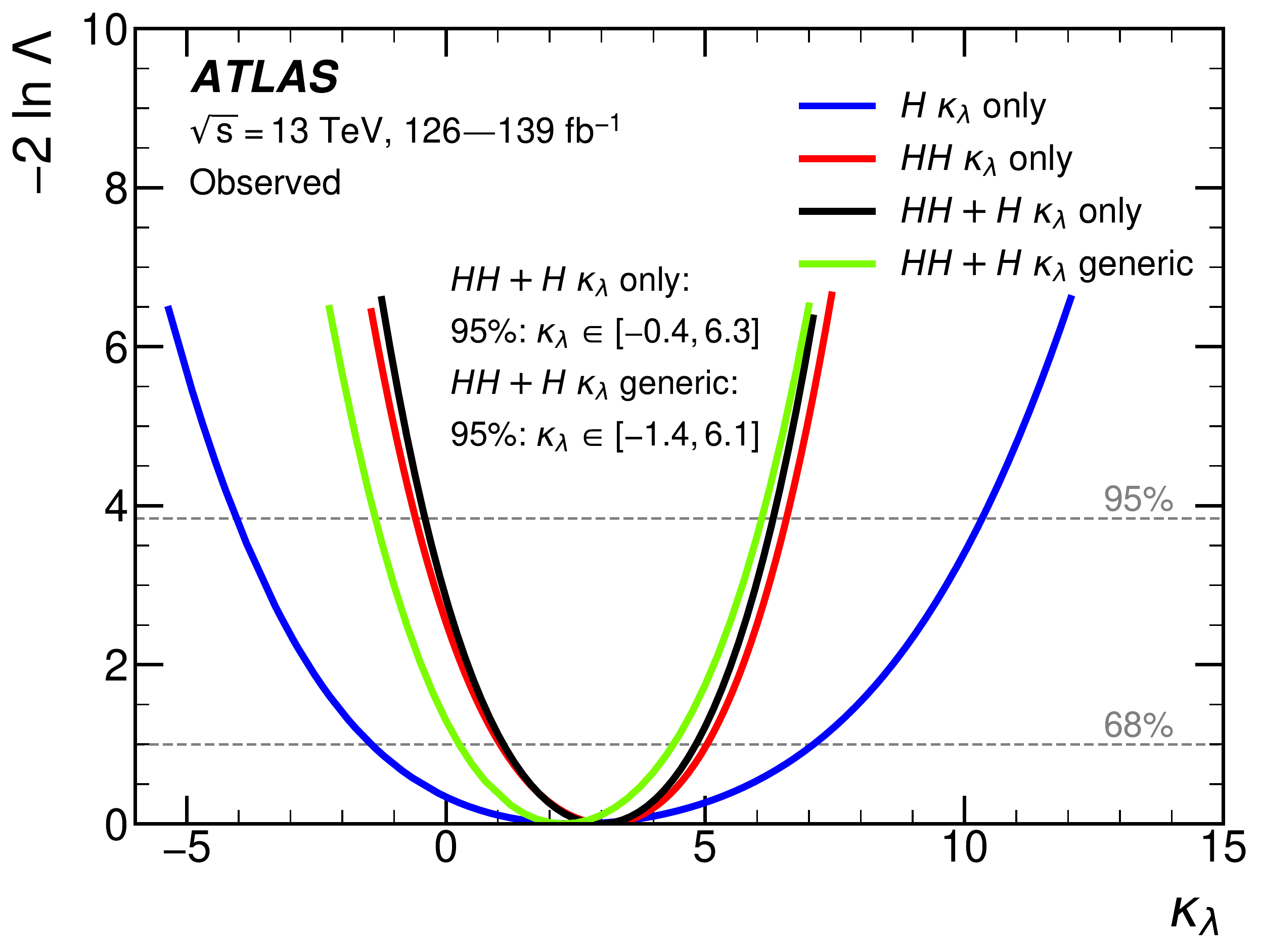}\label{fig:h-hh-klambda-scan_a}}
\subfloat[]{\includegraphics[width=0.5\textwidth,valign=c]{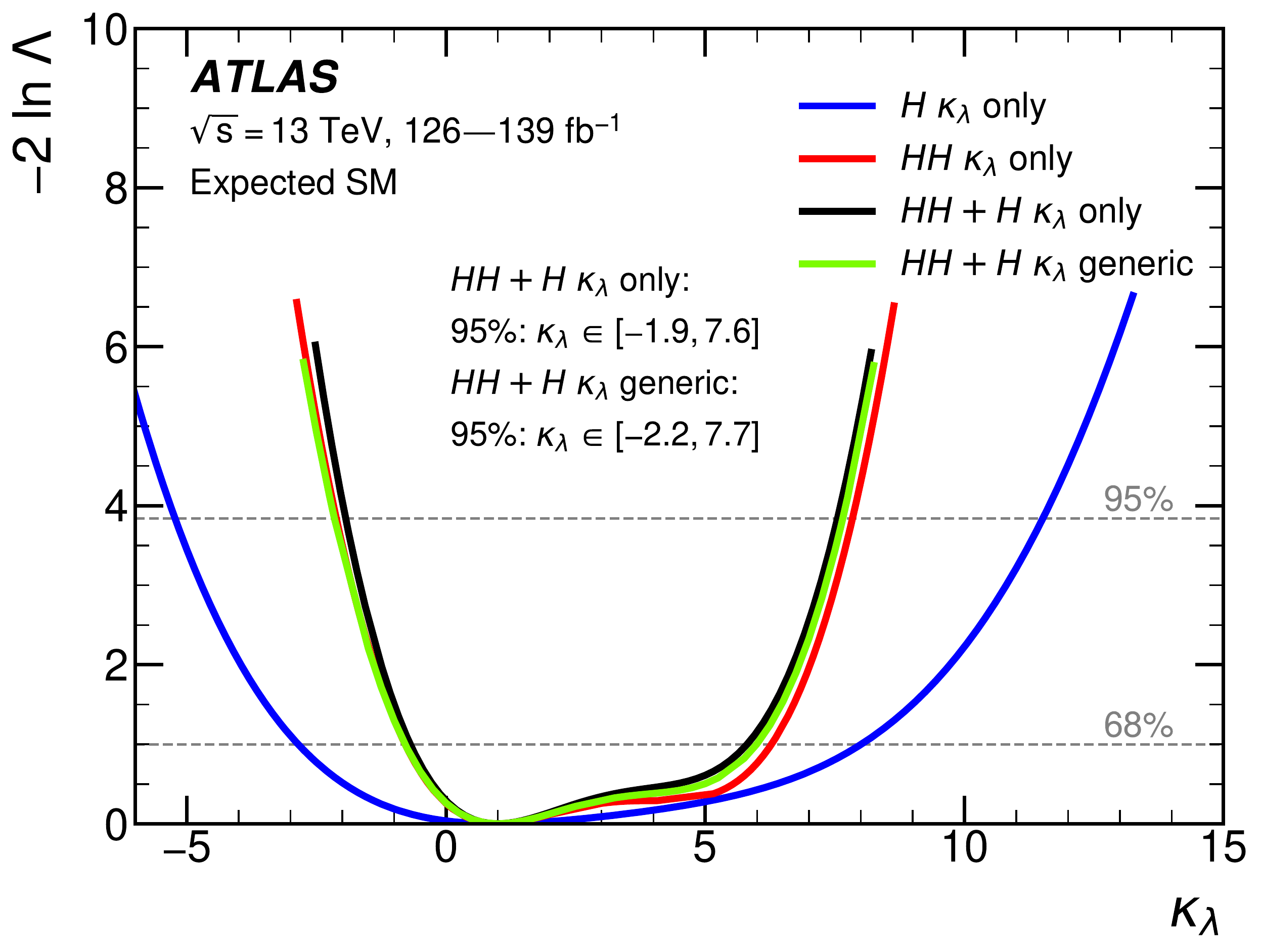}\label{fig:h-hh-klambda-scan_b}}
\caption{Observed (a) and expected (b) values of the test statistic ($-2 \ln{\Lambda}$), as a function  of the \kl\ parameter
for the single-Higgs (blue) and double-Higgs (red) analyses, and their combination (black) derived from the combined single-Higgs and double-Higgs analyses, with
all other coupling modifiers fixed to unity. The combined result for the generic model (free floating \ktop, \kb, \kV\ and \ktau) is also superimposed (green curve).
The observed best-fit value of \kl\ for the generic model is shifted slightly relative to the other models because of its correlation with the best-fit values of
the \kb, \ktop\ and \ktau\ parameters, which are slightly below, but compatible with unity.}\label{fig:h-hh-klambda-scan}
\end{figure}
 
The combined observed (expected) constraints obtained under this hypothesis are $-0.4 < \kl < 6.3$
($-1.9 < \kl < 7.6$) at 95\% CL. All the expected constraints reported in this section are derived from an Asimov dataset generated for
the SM assumption that corresponds to all coupling modifiers equal to unity.
The result is driven by the double-Higgs combination as can be seen in Figure~\ref{fig:h-hh-klambda-scan}.
The expected test statistic ($-2 \ln{\Lambda}$) curve in Figure~\ref{fig:h-hh-klambda-scan}(b) exhibits a
`two-minima-like' structure due to the quadratic dependence of the observed signal yields on the
parameter of interest \kl\ (partially resolved by the $m_{HH}$ kinematic information used in the fit).
The observed curve is more parabolic because the best-fit value of \kl\ is close to the value where the predicted double-Higgs cross-section,
shown in Figure~\ref{fig:hh-xs-exclusion_a}, reaches its minimum.
 
The main advantage of adding the single-Higgs analyses is the possibility of relaxing assumptions about modifiers for couplings to other SM particles.
First, the assumption about the Higgs boson to top-quark coupling modifier, \ktop, can be released.
Thanks to the strong constraints on \ktop\ from the single-Higgs measurements, the constraints on \kl\ obtained from
a fit with a floating value of \ktop\ are almost as strong as those obtained with its value fixed to unity, as reported in Table~\ref{tab:results_kl}.
Two-dimensional contours of $-2 \ln{\Lambda}$ in the \kl--\ktop\ plane are shown in Figure~\ref{fig:h-hh-klambda-kt-scan}.
All other coupling modifiers are fixed to unity in this fit.
 
\begin{figure}[htbp]
\centering
\subfloat[]{\includegraphics[width=0.5\textwidth,valign=c]{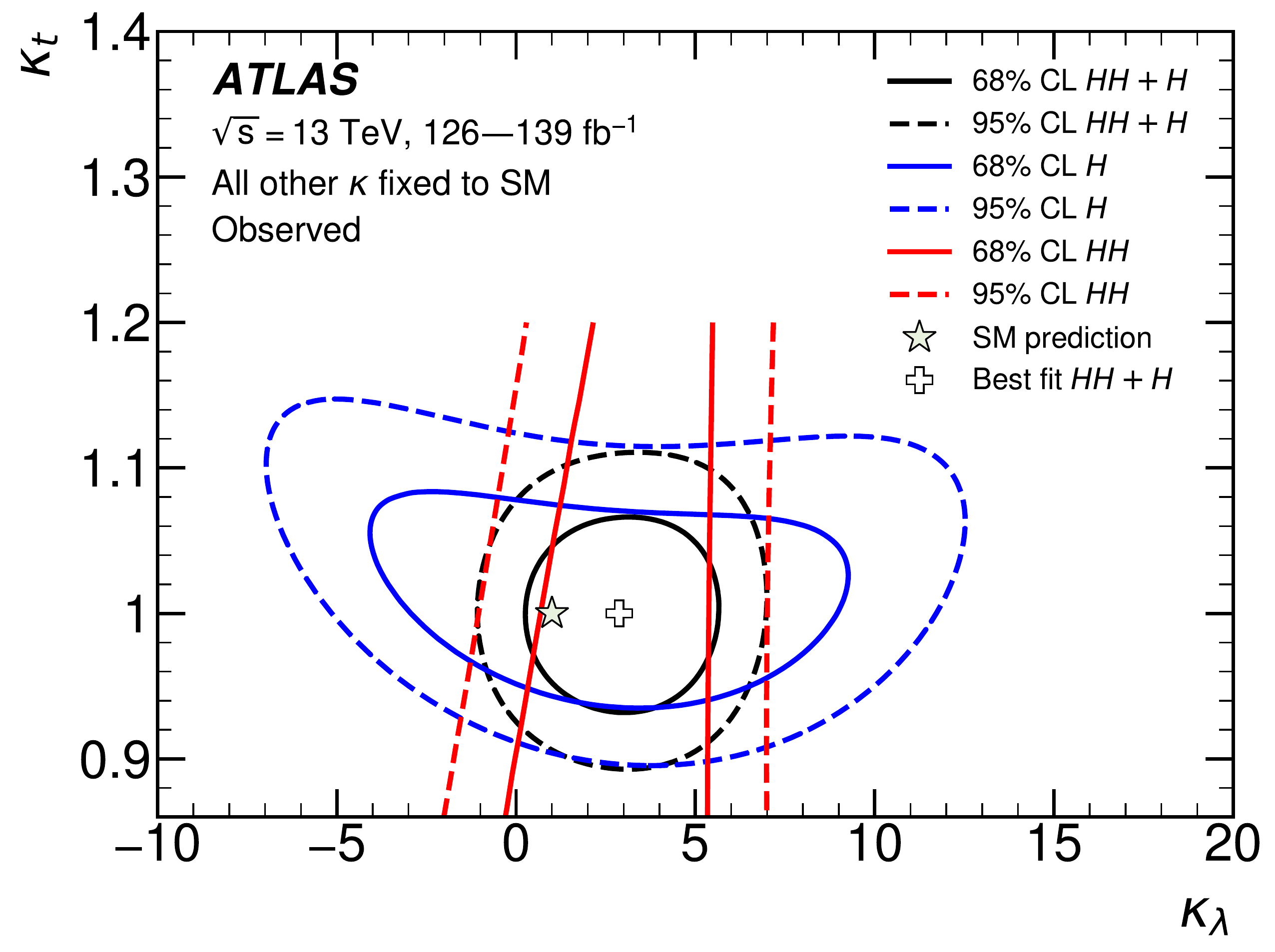}\label{fig:h-hh-klambda-kt-scan_a}}
\subfloat[]{\includegraphics[width=0.5\textwidth,valign=c]{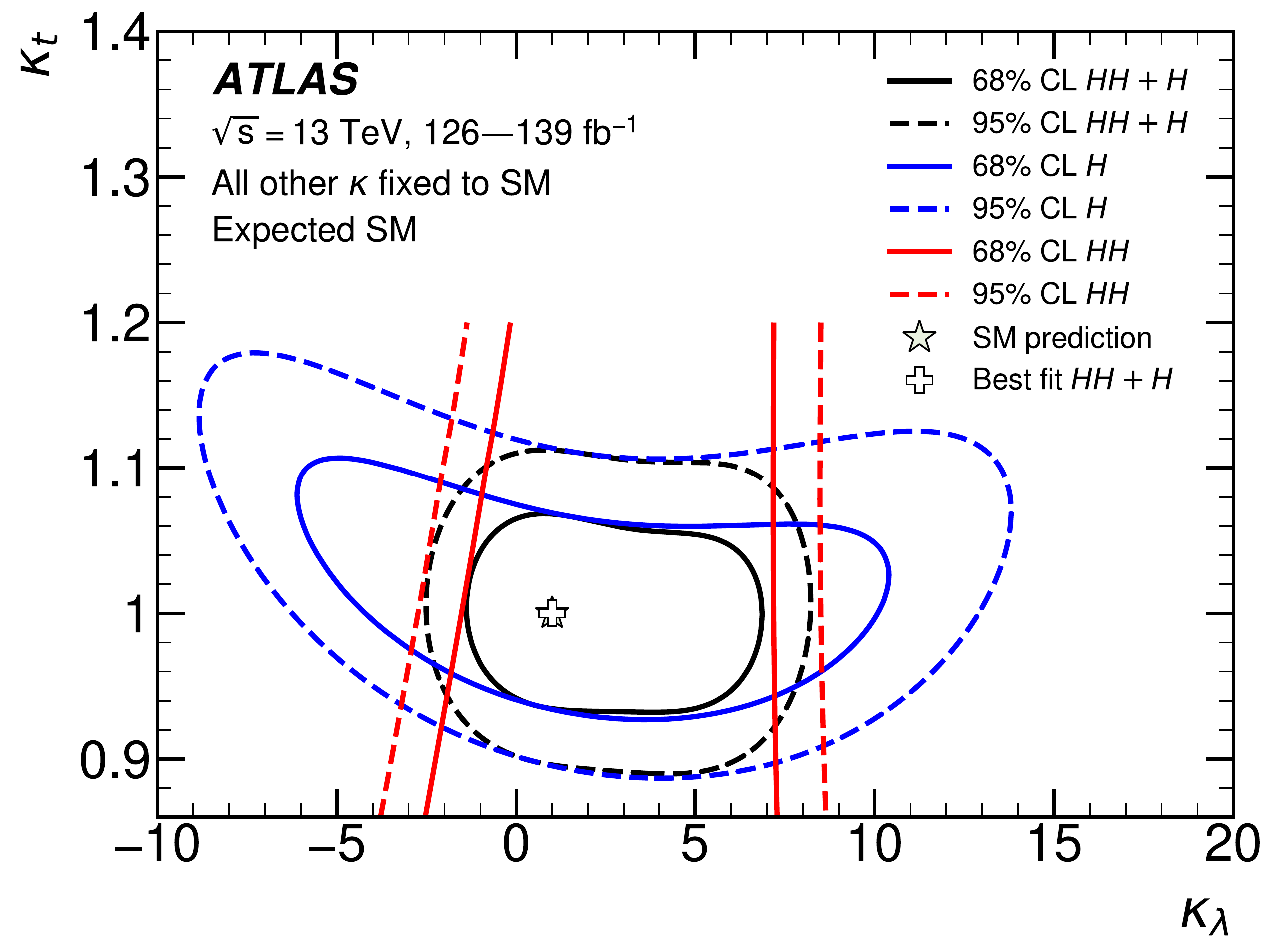}\label{fig:h-hh-klambda-kt-scan_b}}
\caption{Observed (a) and expected (b) constraints in the \kl--\ktop\ plane from  single-Higgs (blue) and double-Higgs (red) analyses, and
their combination (black). The solid (dashed) lines show the 68\% (95\%) CL contours. The double-Higgs contours are shown for values of \ktop\ smaller than 1.2. The observed constraint for the single- and double-Higgs combination for \ktop\ values below unity is slightly less stringent than that for the single-Higgs
fit alone due to the slightly higher best-fit value for this coupling modifier.}
\label{fig:h-hh-klambda-kt-scan}
\end{figure}
 
The most generic model allows all of the coupling modifiers \kl,  $\kappa_t$, $\kappa_b$,
$\kappa_{\tau}$, and $\kappa_V$ implemented in this parameterisation to float freely in the fit. The exception is \ktV, which is fixed to unity since
there is no complete parameterisation of single-Higgs NLO EW corrections as a function of this coupling modifier.
A recent work~\cite{Anisha:2022ctm}, shows that a consistent parameterisation of the $\kappa_V$ and \ktV\
coupling modifiers seems to be possible, though the sensitivity of single-H processes to \ktV\ is shown to be very small.
 
In the combination of the single-Higgs and double-Higgs analyses, an observed (expected) exclusion of
$-1.4 < \kl < 6.1$ ( $-2.2 < \kl < 7.7$) is obtained at 95\% CL in this less model-dependent fit. The values of all the other coupling modifiers
agree with the SM prediction within uncertainties.
The values of the test statistic as a function of \kl\ for this generic model are also shown in Figure~\ref{fig:h-hh-klambda-scan}.
It was checked that for a generic model in which \ktV\ also floats freely in the double-Higgs parameterisation, the observed exclusion constraints on \kl\ weaken by less than 5\%.
In this approach, the $VVHH$ vertex is parameterised in terms of the \ktV\ coupling modifier for the \VBFHH\ process but
the single-Higgs NLO EW corrections are not.
 
\begin{table}[!htbp]
\caption{Summary of \kl\ observed and expected constraints and corresponding observed best-fit values with their uncertainties.
In the first column, the coupling modifiers that are free floating in addition to \kl\ in the corresponding fit are reported.
The uncertainties on \kl\ are extracted from the test statistic curves, which are not expected to follow Gaussian distributions.}
\begin{center}{\def\arraystretch{1.2}\begin{tabular}{lccc}
\toprule
Combination assumption &  Obs.\ 95\% CL  & Exp.\ 95\% CL & Obs.\ value$^{+1\sigma}_{-1\sigma}$ \\
\midrule
$HH$ combination        & $-0.6 < \kl < 6.6$    & $-2.1 < \kl < 7.8$  &  $\kl = 3.1^{+1.9}_{-2.0}$ \\
Single-$H$ combination & ~$-4.0 < \kl < 10.3$   & ~~$-5.2 < \kl < 11.5$ &  $\kl = 2.5^{+4.6}_{-3.9}$ \\
$HH$+$H$ combination      & $-0.4 < \kl < 6.3$    & $-1.9 < \kl < 7.6$  &  $\kl = 3.0^{+1.8}_{-1.9}$ \\
$HH$+$H$ combination, \ktop\ floating                   & $-0.4 < \kl < 6.3$  & $-1.9 < \kl < 7.6$  & $\kl = 3.0^{+1.8}_{-1.9}$\\
$HH$+$H$ combination, \ktop, \kV, \kb, \ktau\ floating  & $-1.4 < \kl < 6.1$  & $-2.2 < \kl < 7.7$  & $\kl = 2.3^{+2.1}_{-2.0}$ \\
\bottomrule
\end{tabular}}
\end{center}
\label{tab:results_kl}
\end{table}
 
\FloatBarrier


\section{Conclusion}
\label{sec:conclusion}

\FloatBarrier
Single- and double-Higgs boson analyses based on the complete LHC Run~2 dataset of 13~\TeV\ proton--proton collisions collected with the ATLAS detector
are combined to investigate the Higgs boson self-interaction and shed more light on the Higgs boson potential,
the source of EW symmetry breaking in the SM.
 
Using the three most sensitive double-Higgs decay channels, $b\bar{b}b\bar{b}$, $b\bar{b}\tau^+\tau^-$ and $b\bar{b} \gamma \gamma$,
an observed (expected) upper limit of 2.4 (2.9) at 95\% CL is set on the double-Higgs signal strength,
defined as the sum of the \ggFHH\ and \VBFHH\ production cross-sections  normalised to its SM prediction.
These processes are directly sensitive to the Higgs boson self-coupling.
This combination can also be used to set a constraint of $-0.6 < \kl < 6.6$ at 95\% CL on the Higgs boson self-coupling
modifier, assuming that the other Higgs boson interactions are as predicted by the SM.
 
Using the \VBFHH\ process, a constraint on the \ktV\ coupling modifier of $0.1 < \ktV < 2.0$ is also derived at 95\% CL,
assuming all other Higgs boson interactions are as predicted by the SM.
 
The measurements from the three double-Higgs decay channels are combined with single-Higgs boson cross-section measurements
from the the $\gamma \gamma$, $ZZ^*$, $WW^*$, $\tau^+ \tau^-$
and $b\bar{b}$ decay channels to derive constraints on \kl\ that are either more stringent or less model-dependent.
Using this combination and assuming that \kl\ is the only source of physics beyond the SM, values of \kl\ outside
the range $-0.4 < \kl < 6.3$ are excluded at 95\% CL, with an expected excluded range of  $-1.9 < \kl < 7.6$.
If assumptions about the other coupling modifiers, \ktop, \kb\ \ktau, and \kV, are relaxed, this constraint becomes $-1.4 < \kl < 6.1$ at 95\% CL,
where the expected interval under the SM assumption is  $-2.2 < \kl < 7.7$. This constraint on the Higgs boson self-coupling is
not quite as strong but less model-dependent.
This study provides the most stringent constraints on Higgs boson self-interactions to date.


\section*{Acknowledgements}


We thank CERN for the very successful operation of the LHC, as well as the
support staff from our institutions without whom ATLAS could not be
operated efficiently.
 
We acknowledge the support of
ANPCyT, Argentina;
YerPhI, Armenia;
ARC, Australia;
BMWFW and FWF, Austria;
ANAS, Azerbaijan;
CNPq and FAPESP, Brazil;
NSERC, NRC and CFI, Canada;
CERN;
ANID, Chile;
CAS, MOST and NSFC, China;
Minciencias, Colombia;
MEYS CR, Czech Republic;
DNRF and DNSRC, Denmark;
IN2P3-CNRS and CEA-DRF/IRFU, France;
SRNSFG, Georgia;
BMBF, HGF and MPG, Germany;
GSRI, Greece;
RGC and Hong Kong SAR, China;
ISF and Benoziyo Center, Israel;
INFN, Italy;
MEXT and JSPS, Japan;
CNRST, Morocco;
NWO, Netherlands;
RCN, Norway;
MEiN, Poland;
FCT, Portugal;
MNE/IFA, Romania;
MESTD, Serbia;
MSSR, Slovakia;
ARRS and MIZ\v{S}, Slovenia;
DSI/NRF, South Africa;
MICINN, Spain;
SRC and Wallenberg Foundation, Sweden;
SERI, SNSF and Cantons of Bern and Geneva, Switzerland;
MOST, Taiwan;
TENMAK, T\"urkiye;
STFC, United Kingdom;
DOE and NSF, United States of America.
In addition, individual groups and members have received support from
BCKDF, CANARIE, Compute Canada and CRC, Canada;
PRIMUS 21/SCI/017 and UNCE SCI/013, Czech Republic;
COST, ERC, ERDF, Horizon 2020 and Marie Sk{\l}odowska-Curie Actions, European Union;
Investissements d'Avenir Labex, Investissements d'Avenir Idex and ANR, France;
DFG and AvH Foundation, Germany;
Herakleitos, Thales and Aristeia programmes co-financed by EU-ESF and the Greek NSRF, Greece;
BSF-NSF and MINERVA, Israel;
Norwegian Financial Mechanism 2014-2021, Norway;
NCN and NAWA, Poland;
La Caixa Banking Foundation, CERCA Programme Generalitat de Catalunya and PROMETEO and GenT Programmes Generalitat Valenciana, Spain;
G\"{o}ran Gustafssons Stiftelse, Sweden;
The Royal Society and Leverhulme Trust, United Kingdom.
 
The crucial computing support from all WLCG partners is acknowledged gratefully, in particular from CERN, the ATLAS Tier-1 facilities at TRIUMF (Canada), NDGF (Denmark, Norway, Sweden), CC-IN2P3 (France), KIT/GridKA (Germany), INFN-CNAF (Italy), NL-T1 (Netherlands), PIC (Spain), ASGC (Taiwan), RAL (UK) and BNL (USA), the Tier-2 facilities worldwide and large non-WLCG resource providers. Major contributors of computing resources are listed in Ref.~\cite{ATL-SOFT-PUB-2021-003}.


\printbibliography
 
\clearpage

\clearpage
 
\end{document}